\documentclass[10pt,aps,prl,twocolumn,english]{revtex4-1}

\usepackage{graphicx}
\usepackage{amsmath}
\usepackage{amssymb}
\usepackage{dsfont}
\usepackage{fancyhdr}
\usepackage{yfonts}
\usepackage{braket}
\usepackage{siunitx}
\pdfoutput=1
\newcommand{\eg}{e.g.~}
\newcommand{\ie}{i.e.~}
\usepackage[utf8]{inputenc}
\DeclareFontFamily{U}{BOONDOX-calo}{\skewchar\font=45 }
\DeclareFontShape{U}{BOONDOX-calo}{m}{n}{
  <-> s*[1.05] BOONDOX-r-calo}{}
\DeclareFontShape{U}{BOONDOX-calo}{b}{n}{
  <-> s*[1.05] BOONDOX-b-calo}{}
\DeclareMathAlphabet{\mathcalboondox}{U}{BOONDOX-calo}{m}{n}
\SetMathAlphabet{\mathcalboondox}{bold}{U}{BOONDOX-calo}{b}{n}
\DeclareMathAlphabet{\mathbcalboondox}{U}{BOONDOX-calo}{b}{n}

\usepackage{hyperref}
\hypersetup{
  colorlinks   = true, 
  urlcolor     = blue, 
  linkcolor    = blue,
  citecolor   = blue,
}

\begin{document}
\title{Strong Purcell effect on a neutral atom trapped in an open fiber cavity}
\author{J. Gallego}\email[]{gallego@iap.uni-bonn.de}
\author{W. Alt}
\author{T. Macha}
\author{M. Martinez-Dorantes}
\author{D. Pandey}
\author{D. Meschede}
\affiliation{Institut f\"ur Angewandte Physik der Universit\"at Bonn, Wegelerstrasse 8, D-53115 Bonn, Germany}


\begin{abstract}  
We observe a sixfold Purcell broadening of the D$_2$ line of an optically trapped
$^{87}\text{Rb}$ atom strongly coupled to a fiber
cavity. Under external illumination by a near-resonant laser, up to $90\%$ of the atom's fluorescence is emitted into
the resonant cavity mode.
The sub-Poissonian statistics of the cavity output and the Purcell enhancement of the atomic decay rate are confirmed by the observation of a strongly 
narrowed antibunching dip in the 
photon autocorrelation function.
The photon leakage
through the higher-transmission mirror of the single-sided resonator is the dominant contribution to the
field decay ($\kappa\!\approx\!2\pi\!\times\!50\,\text{MHz}$),
thus offering a high-bandwidth,
fiber-coupled channel for photonic interfaces such as 
quantum memories and single-photon sources. 
\end{abstract}
\maketitle
  The modification of the photoemission properties of matter is a field of broad and current interest
  in the pursuit of controlled and efficient light--matter interfaces. The use of a resonator to enhance the
  spontaneous emission rate of an atom was
  proposed by E.M. Purcell~\cite{Purcell1946} and later observed in the
  microwave~\cite{Goy1983} and optical domains~\cite{Heinzen1987}. According to his proposal, in the presence of
  a cavity with a single resonant field mode, the decay rate ($2\gamma$) of an excited atom is
  enhanced by the factor $f_\text{P}$, which in
  cavity quantum electrodynamics (CQED) is related to the cooperativity of the system by
  $f_\text{P}\!=\!2C\!=\!g^2/(\kappa\gamma)$.
  Here, $g$ is the 
  atom--field interaction strength, and $\kappa$ 
  stands for the decay rate of the cavity field.
  Notably, the directionality of the atomic emission is also enhanced,
  and the cavity mode collects
  a fraction of the photons given by $f_\text{P}/(f_\text{P}\!+\!1)$. Both effects facilitate 
  the generation and efficient collection of single photons and, thus, 
  the study of the Purcell effect has been extended to multiple
  types of emitters. These include atoms~\citep{Lien2016,Ballance2017},
  quantum dots~\cite{Gerard1998,Laucht2012},
  and a variety of other solid-state
  systems~\cite{Bienfait2016,Sumikura2016,Ding2016,Dibos2017} amongst others,
  with particular 
  emphasis on the development of single-photon sources
  (SPS)~\cite{McKeever1992,Kuhn2002,Press2007,Kang2011,Mucke2013,Aharonovich2016,Dolan2018}.
  The required high cooperativities can be
  obtained by tailoring the resonator, \ie reducing the mirror transmission and losses
  (for low $\kappa$), or the
  mode volume (for high $g$).
  However, to build an efficient SPS with a high-rate output, the resonator must supply a photon out-coupling rate
  faster than the spontaneous emission rate, \ie $\kappa\!\gg\!\gamma$.
  A strong Purcell enhancement in such \textit{open}-cavity
  platforms can only be realized by using microresonators with reduced mode volume~\citep{Vahala2003} or 
  emitters with 
  narrow emission lines, like neutral atoms and ions~\citep{Colombe2007,Steiner2013,Sayrin2015}. Large Purcell
  factors have been recently realized for narrow forbidden lines of
  $\text{Er}^{3+}$ ions in solid-state
  hosts~\citep{Dibos2017}. However, these solid-state systems are still not within the desired regime
  of $g\!>\!\kappa\!\gg\!\gamma$, where the dominant coherent interaction allows for the reversible storage
  of quantum information, of particular interest for the creation of efficient hybrid quantum communication links.
  
  Atom-based platforms guarantee highly controllable,
  coherent, and reproducible photon sources, without the need for cryogenic equipment.
  Of particular interest is their integration with fiber Fabry-P\'erot cavities (FFPCs)~\cite{Hunger2010},
  which can host mode volumes of the order of the emission wavelength $\lambda^3$~\cite{Kaupp2016} while
  offering high tunability, intrinsic fiber coupling, and a good radial optical access, which is
  critical for the external manipulation of the emitter.
  The rapid progress seen in FFPC platforms
  in recent years~\cite{Muller2009,Flowers2012,Albrecht2014,Jeantet2016}
  includes the demonstration of Purcell broadening on the photoemission of
  atoms~\cite{Lien2016,Ballance2017,Takahashi2017}; however, in all reported cases,
  single-atom Purcell factors
  remained well below unity. Substantial cooperativities have only been shown
  for transient dense clouds of
  atoms~\cite{Lien2016}, extremely \textit{closed} macroscopic cavities
  ($\kappa\!\approx\!\gamma$)~\cite{McKeever1992,Mucke2013},
  or in a setup without external addressing~\citep{Colombe2007} -- all those scenarios
  preventing its use as a high-bandwidth SPS.
  In this letter, we demonstrate the highest Purcell broadening so far directly observed for an externally driven,
  single atom.
  Our system consists of a neutral $^{87}\text{Rb}$ atom optically trapped inside
  an open FFPC. We show that the platform operates in the desired open cavity regime, with an
  emission spectrum displaying a line broadening corresponding
  to $f_\text{P}\!\approx\!5$, a factor that increases up to 20 when better pumping the atom towards the
  strongest transition.
  The homogeneous nature of the broadening is confirmed by analyzing the output of the cavity with a 
  Hanbury Brown--Twiss setup, which
  reveals a narrow antibunching dip corresponding to the generation of single photons from an atom with a
  strongly shortened lifetime.

\begin{figure}
    	\includegraphics[width=1\columnwidth]{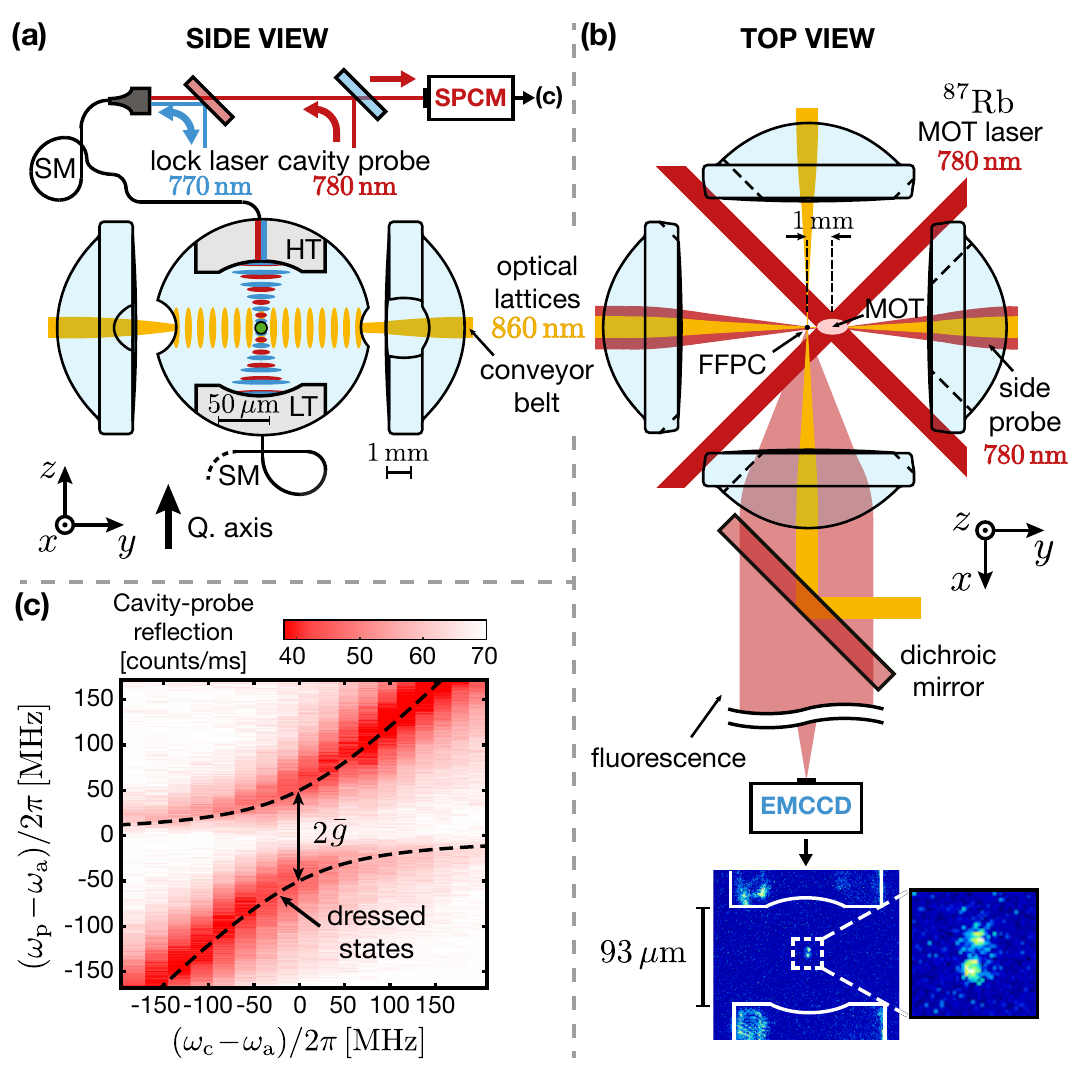}
  		\caption{Experimental setup.
       (a) Side- and (b) top-view technical drawing of the main
        components and the relevant light fields, showing the cavity- and 
        side-probe configurations, respectively. The 
        bottom inset shows a fluorescence image of two atoms coupled to the resonator ($20\,\text{ms}$ exposure
        time).
        (c) Probe power reflected from the cavity ($\kappa\!=\!2\pi\times 58\,\text{MHz}$),
        showing the energy bands of the coupled atom-cavity system
        for a single atom pumped well below the saturation photon number. The dashed lines display the 
        dressed states of the addressed cycling transition
        ($\ket{F,m_F}\!=\!\ket{2,2}\!\leftrightarrow\!\ket{3^\prime,3^\prime}$).
        For each cavity frequency ($\omega_\text{c}$,
        horizontal axis),
        the probe frequency $\omega_\text{p}$ is
        scanned  through resonance.}
        {\label{fig:Setup}}
	\end{figure}

	The core of our experimental apparatus is a one-sided high-finesse FFPC
    (described in detail in~\citep{Gallego2016}) where one 
    of the fiber mirrors features a relatively high transmission (HT),
    thus serving as an efficient input-output coupling
    channel. The coating characteristics and the 
    length of the resonator lead to
    an initial field decay rate of $\kappa\!\approx\!2\pi\!\times\!25\,\text{MHz}$, which later degraded under
    vacuum conditions (see Supplemental Material).
    The cavity is placed
    at the center of an integrated, compact mount featuring four
    aspheric lenses with high numerical aperture ($\text{NA}\!=\!0.5$, see Fig.~\ref{fig:Setup}),
    which are the keystone of the high degree
    of optical control in our system~\cite{Dorantes2017}.
    Amongst other applications, they are used to strongly focus two pairs of counter-propagating,
    red-detuned dipole-trap beams ($860\,\text{nm}$), which create a 2D optical lattice
    (see Figs.~\ref{fig:Setup}(a,b)).
    The ``lock laser'' ($770\,\text{nm}$) employed to stabilize the resonator's length~\citep{Gallego2016}
    also serves as a blue-detuned intracavity dipole trap,
    resulting in a 3D
    subwavelength confinement of the atom at the antinodes of the cavity mode resonant with the
    $\text{D}_2$ line of $^{87}\text{Rb}$ ($780\,\text{nm}$). 

    In a typical experiment, a few tens
    of neutral $^{87}\text{Rb}$ atoms are
    trapped from the background gas ($10^{-9}\,\text{mbar}$) and cooled down
    to $\sim\!50\,\mu\text{K}$
    by a magneto-optical trap (MOT) positioned
    $1\,\text{mm}$ away from the cavity center, as depicted in Fig.~\ref{fig:Setup}(b).
    The atoms are subsequently loaded into one of the optical lattices -- which acts as an optical
    conveyor belt~\cite{Schrader2001} -- and transported into the
    cavity region. Once inside the resonator,
    the presence of a coupled atom is detected by its interaction with a near-resonant probe field
    ($780\,\text{nm}$), which is either injected into the cavity or focused on the atom by the in-vacuum lenses
    (Fig.~\ref{fig:Setup}(a) and~\ref{fig:Setup}(b), respectively).

    In the first case, referred to as ``cavity probe'',
    the presence of an atom
    manifests as a rise in the reflected probe power when the field and the cavity are resonant with the atom.
    The resulting increase in counts -- detected by a
    single-photon counter module (SPCM) -- is used for real-time feedback to halt the 
    transport mechanism as soon as an atom strongly couples to the cavity mode.
    This ensures the coupling of a single atom in $\sim\!85\%$ of the cases.
    By scanning both the probe frequency and the cavity resonance, a clear avoided
    crossing appears between
    the dressed states of the coupled system, see Fig.~\ref{fig:Setup}(c).
    From the vacuum Rabi splitting we obtain an
    average coupling constant of
    $\bar{g}\!=\!2\pi\times(49.94\pm 0.12)\,\text{MHz}$ following a Gaussian distribution of width 
    $\sigma_g\!=\!2\pi\times(18.2\pm0.2)\,\text{MHz}$, as a result of different positions of the atom
    inside the cavity mode.
    The system's average single-atom cooperativity is thus approximately $\bar{C}\!\approx\!7.2$
    (distributed
    over a range $C\!\in[2.9,13.4]$), which demonstrates the high cooperativity of our platform even
    for bandwidths $\kappa\gg\gamma$. Such a fast coherent interaction is an essential prerequisite for
    reversible processes required for photon storage and retrieval~\cite{Gorshkov2007}.

    For applications such as SPS, the atom must be directly addressed by an external driving field.
    This is the role of the second type of illumination --
    depicted in Fig.~\ref{fig:Setup}(b) as ``side-probe'' -- where the atoms are continuously driven by
    (typically red-detuned) light
    in a lin$\perp$lin configuration. The resulting polarization gradient
    provides one-dimensional cooling and prevents the formation of intensity standing waves.
    The in-vacuum lenses enable not only the required strong focusing of the illumination beam, 
    but also the efficient collection of the photons scattered into free space by the atoms. These
    are imaged onto
    an electron-multiplying CCD (EMCCD) camera that yields high-resolution fluorescence images,
    critical for the
    estimation of the position
    and number of atoms (inset of Fig.~\ref{fig:Setup}(b)). The large fraction of 
    light scattered into the cavity is monitored by the SPCM and, along with the 
    camera counts, it provides the necessary information to characterize the system's
    photoemission properties. Such a study requires a 
	model for the scattering rates under external driving of the atom. As described
    in the Supplemental Material, for a continuous driving field of Rabi
	frequency $\Omega$ and frequency $\omega_\text{p}$, the rate of photons emitted in
    free space ($R_\text{f-s}$) or into the cavity ($R_\text{c}$) are
	\begin{subequations}\label{eq:ScatteringRates}
	\begin{align}
		R_\text{f-s} &= \frac{\Omega^2/(2\gamma)}
    	{1+\Delta_\text{a}^2/\gamma^2}\,\frac{1}{|1+2\tilde{C}|^2} \label{eq:ScatteringRateAtom}\\
		R_\text{c} &= \frac{\Omega^2}{\gamma\,C}\,\frac{|\tilde{C}|^2}{|1+2\tilde{C}|^2}\,
        ,\label{eq:ScatteringRateCav}
	\end{align}
	\end{subequations}
	where we have introduced the complex cooperativity parameter (see e.g.~\cite{Murr2003})
	\begin{equation}\label{eq:ComplexCooperativity}
		\tilde{C} =\frac{g^2}{2(\kappa-i\Delta_\text{c})(\gamma-i\Delta_\text{a})}\, ,
	\end{equation}
	with $\Delta_\text{c/a}\!=\!\omega_\text{p}-\!\omega_\text{c/a}$. Here $\omega_\text{c}$ and
    $\omega_\text{a}$ correspond
    to the cavity resonance and the ac-Stark shifted frequency of the atomic cycling transition,
    respectively.

    The model assumes that the atom is a weakly driven two-level system. 
    We consider effects due to power saturation of multilevel transitions
    by comparing the model to numerical simulations based on the master equation formalism.
    \begin{figure}
		\includegraphics[width=1\columnwidth]{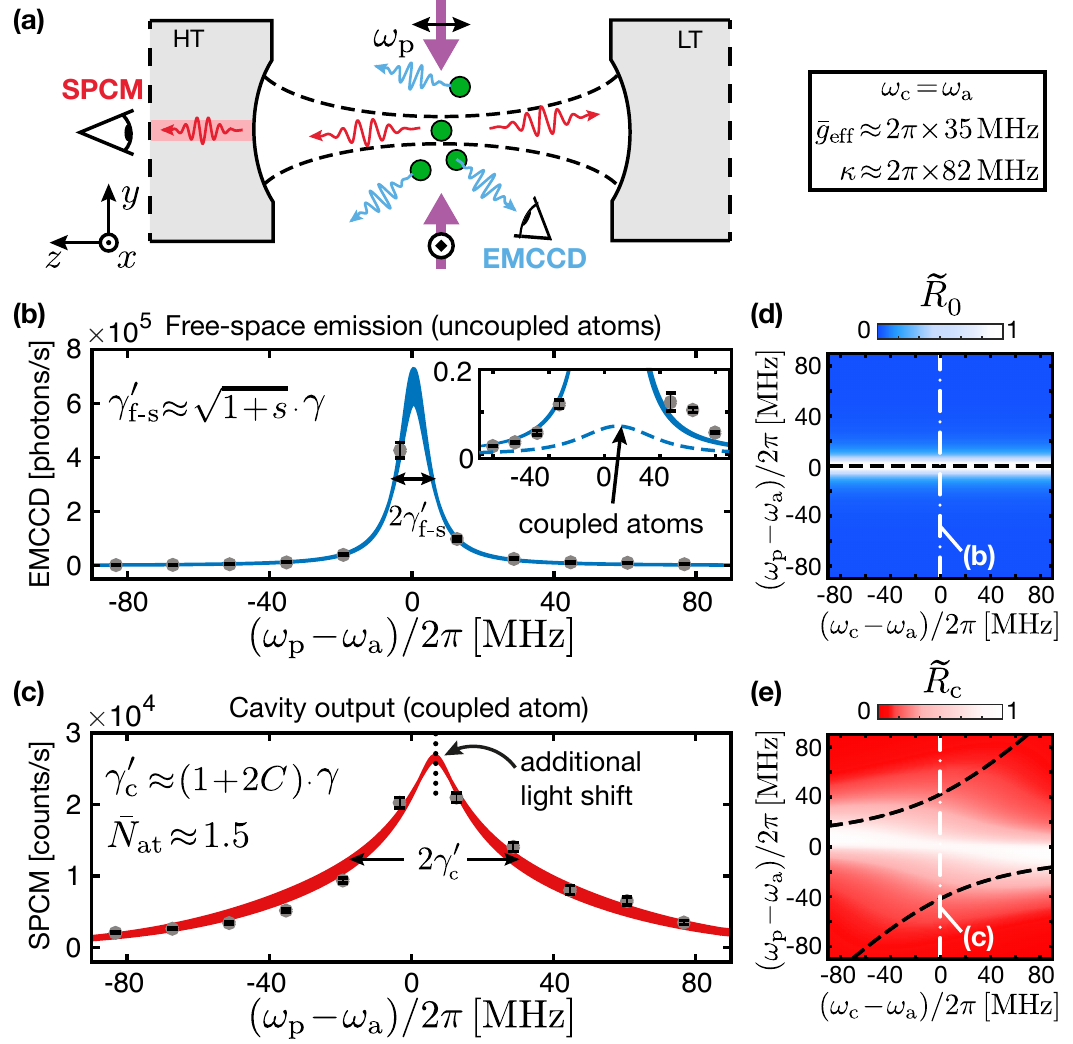}
 		\caption{Purcell-induced broadening of the atomic line shape. (a) Simplified
        sketch of the measurement for coupled and uncoupled atoms. The lin$\perp$lin
        probe light (purple) is
        scattered either into the cavity mode (red arrows) or into free space (blue arrows).
        (b) Free-space emission spectrum of atoms trapped outside the cavity mode, error bars
        indicate the one-sigma statistical error of the mean.
        The solid line is a fit to a Lorentzian curve, while
        the dashed line in the inset shows the negligible
        free-space contribution expected from
        atoms coupled to the cavity. (c) Emission line shape of the atom-cavity
        system displaying a clear Purcell 
        broadening. The solid line represents a fit to a
    	convolution of a Lorentzian curve of half-width
    	$\gamma^\prime_\text{c}\!=\!(1+g_\text{eff}^2/\kappa\gamma)\cdot\gamma$ and a Gaussian
    	distribution of coupling strengths
    	with mean $\bar{g}_\text{eff}$ and variance $\sigma_g$ (see main text).
        Differences in $m_F$-states population lead to
        an additional light shift.
        (d) and (e) are theoretical emission rates for an uncoupled atom in 
        free space ($\tilde{R}_0$, blue) and the
        cavity output of a coupled system ($\tilde{R}_\text{c}$, red, $N_\text{at}\!=\!1$)
        using the fit parameters, both normalized to their maximum value.
        The dashed, black lines represent the
        eigenenergies of the system.}{\label{fig:Broadening}}
	\end{figure}
	Externally driving the atom with near-resonant lin$\perp$lin probe light
    leads to a reduced effective
    coupling strength $g_\text{eff}$ due to the addressing of transitions weaker than the strongest one.
	In this case $\kappa\!>\!g_\text{eff}\!\gg\!\gamma$, and the system enters the \textit{fast-cavity regime} 
    where the resonator's
    output ($\propto\!R_\text{c}$) corresponds to
	a broadened Lorentzian curve of half-width $\gamma^\prime_\text{c}\!=\!(1+2C)\cdot\gamma$, which
    is a direct signature of the 
	Purcell-enhanced atomic decay rate. 
	The effect of the resonator becomes evident when comparing the cavity output line shape to that
    of uncoupled atoms, given by
    $R_0\!\!:=\!R_\text{f-s}(g\!=\!0)$.
	Both emission spectra are measured by loading an atomic 
    ensemble into the
    cavity region, with $\bar{N}_\text{at}\!=\!1.5$ atoms coupling to the resonator on average. 
	The ensemble is illuminated with side-probe light, the frequency of which is
    scanned through the system's resonance
    (with $\omega_\text{a}\!=\!\omega_\text{c}$, see Fig.~\ref{fig:Broadening}(a)). The resulting
    scattering rates are estimated from the detected photon counts as described in the Supplemental Material.
    
    Figs.~\ref{fig:Broadening}(b,d) display the free-space emission line shape of the system, estimated from
    the camera counts. We assume that such a spectrum corresponds exclusively to uncoupled atoms,
    since the 
    free-space contribution from coupled atoms is negligible given that
    $R_\text{f-s}/R_0\!=\!|1+2C|^{-2}\!\approx\!1.5\,\%$
    (see inset in Fig.~\ref{fig:Broadening}(b)). The resonance follows a Lorentzian curve of half-width
    $\gamma^\prime_\text{f-s}\!\approx\!1.65\cdot\gamma$ attributed to a combination of power broadening
    and inhomogeneous ac-Stark shifts of the ensemble
    in the outer regions of the 3D lattice. We can neglect Doppler and collisional
    broadenings due to the individual, tight confinement of the 
    atoms.
    On the other hand, the line shape of the cavity output, shown in Figs.~\ref{fig:Broadening}(c,e),
    stems from
    photons scattered solely by coupled atoms,
    and corresponds to the Purcell-broadened spectrum of the coupled system.
    The effect of the cavity on the atomic properties manifests as a clear broadening of
    $\gamma^\prime_\text{c}/\gamma\!\approx\!8.4$
    (corresponding to $\gamma^\prime_\text{c}/\gamma\!\approx\!5.9$ for $N_\text{at}\!=\!1$),
    which yields an average single-atom cooperativity of $\bar{C}\!=\!2.5\pm 0.3$.
    We can exclude other effects
    such as power broadening (as the saturation parameter $s$ scales with $(1\!+\!2C)^{-2}$) or 
    inhomogeneous light shifts (since the coupled atoms are confined in the well-defined central
    region of the 3D trap).
    The difference to the value
    estimated from Fig.~\ref{fig:Setup} comes from higher mirror losses
    and a different $m_F$-level distribution (estimated
    to lead to $g_\text{eff}\!\approx\!0.73\!\cdot\!\,g$).
    Such a single-atom cooperativity surpasses by more than an order of magnitude those of
    similar reported fiber-cavity
    systems with externally driven
    atoms~\cite{Lien2016,Ballance2017,Takahashi2017}, and could be further increased
    by precise positioning of the atom and a far-detuned driving that addresses stronger transitions
    (see Supplemental Material for a discussion on the effective coupling strength).

\begin{figure}
 		\includegraphics[width=1\columnwidth]{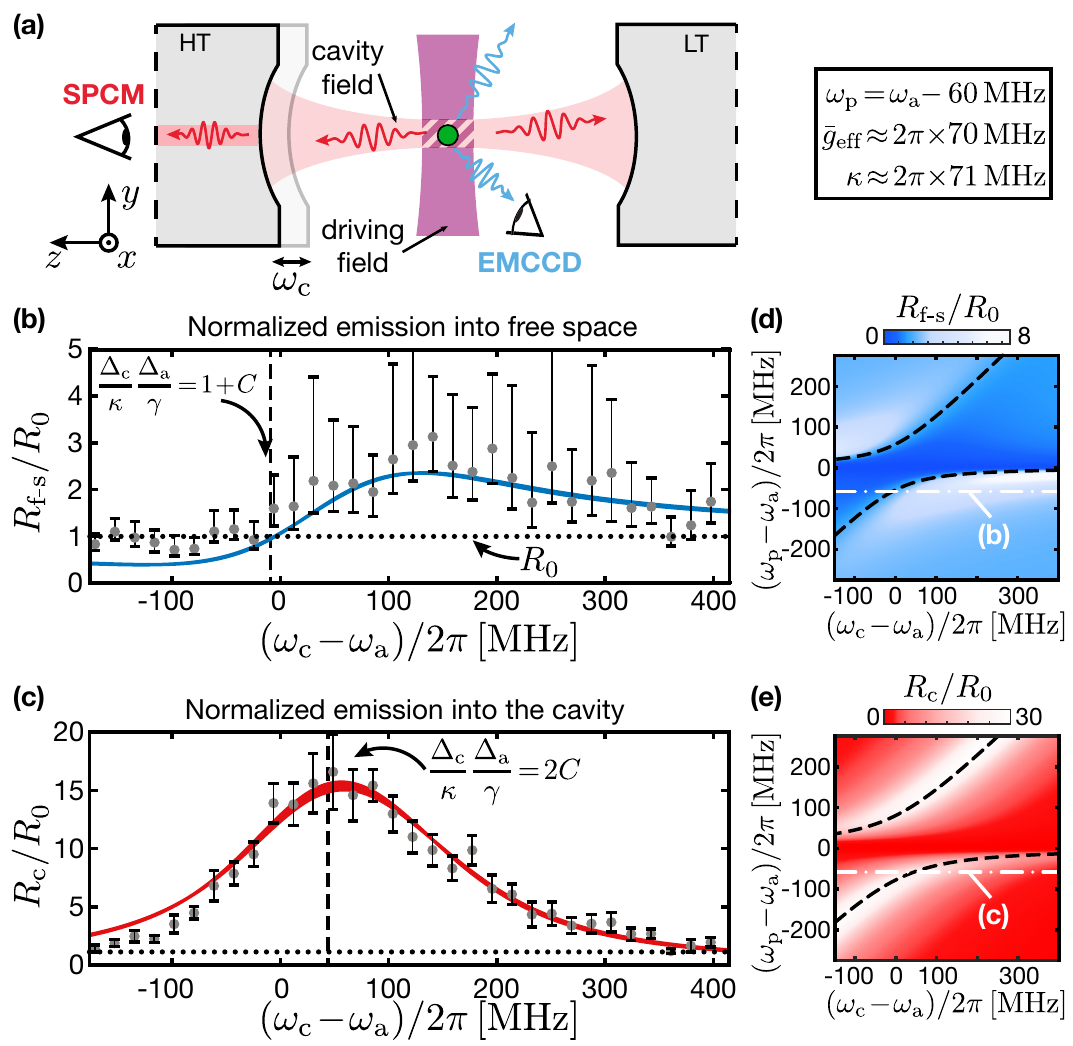}
 		\caption{Cavity backaction on the atomic photoemission rate. (a) Simplified
        experimental scheme depicting the two (interfering) fields.
        (b) and (c)
        display the system's emission in free space and into the cavity respectively,
        normalized to that of an uncoupled atom (dotted black line).
        The error bars are extracted from data bootstrapping 
        and Monte Carlo error propagation. The uncertainties in 
		(b) are a consequence of the small free-space collection efficiency and the short trapping lifetimes.
        The solid lines correspond to the fit to an expanded version of Eq.~\ref{eq:ScatteringRateAtom}
        (blue) and~\ref{eq:ScatteringRateCav} (red)
        including a convolution for different coupling strengths and optical pumping effects estimated from the
		master equation simulation (see the Supplemental Material).
	 	(d) and (e) show the corresponding full spectra for the normalized emission
 		into both free space and the cavity. The black dashed lines represent
		the $\Delta_\text{c}\Delta_\text{a}/(\kappa\gamma)\!=\!1\!+\!C$ region in (d), and
		the dressed eigenbands $\Delta_\text{c}\Delta_\text{a}/(\kappa\gamma)\!=\!2C$
        in (e).}{\label{fig:CavityBackaction}}
	\end{figure}

	The faster atomic decay inferred from the broadening
    is also associated to a strong directionality of the photoemission
    into the cavity mode, given by the ratio
	\begin{equation}\label{eq:RatesRatio}
		R_\text{c} / R_\text{f-s} =\frac{2C}{1+\Delta_\text{c}^2/\kappa^2}\, ,
	\end{equation}
	where the denominator indicates the reduction of the Purcell effect for an off-resonant cavity.
    Such a directionality is integral to the development of efficient SPS, and its consequences 
    become apparent in our system, due to the continuous
    driving of the atom inside a high-finesse 
	cavity. In such a scenario, the high fraction of photons scattered into the resonator
    build up a cavity field that can reach an amplitude
	comparable to that of the side-probe driving field (or even the same amplitude for
    $\kappa\!\rightarrow\!0$~\cite{Alsing1992}).
    The interference between both fields alters the total effective driving
    experienced by the atom and, therefore, 
    its total scattering rate.
    The effect depends on the relative phase between both fields, given by the 
    phaseshifts acquired from the
    off-resonant atomic scattering $\propto\!\Delta_{a}/\gamma$
    and the cavity roundtrips $\propto\!\Delta_\text{c}/\kappa$ (see e.g.~\cite{Ballance2017}).
    Such effect, dubbed cavity
    backaction~\cite{Alsing1992,Tanji2011,Reimann2015,Ballance2017},
    is fully contained in the simplified quantum model of Eqs.~\ref{eq:ScatteringRates}, and 
    confirmed in our experiment by characterizing the photoemission of a single atom
    coupled to the resonator under red-detuned illumination (see Fig.~\ref{fig:CavityBackaction}).

    Figs.~\ref{fig:CavityBackaction}(b,d) show how the feedback of the cavity field on the atomic emission
    leads to either a reduction or an enhancement of photon scattering into free space,
    depending on the cavity resonance frequency.
    The curve follows the
	behavior predicted by 
	Eq.~\ref{eq:ScatteringRateAtom}, including the detuning conditions that lead to the boundary
    $R_\text{f-s}\!=\!R_0$ between
	scattering enhancement and reduction (given by
    $\Delta_\text{a}/\gamma\cdot\Delta_\text{c}/\kappa\!=\!1\!+\!C$).
    The corresponding cavity emission (shown in Figs.~\ref{fig:CavityBackaction}(c,e)) displays a
	curve with a maximum emission rate 15 times higher than the total emission from an
    uncoupled atom. The peak corresponds to the probe light being resonant to one of the dressed
    states of the coupled system, and results from the combination of
    constructive backaction and Purcell directionality, with an
	estimated average cooperativity of $\bar{C}\!=\!11.3\pm 1.0$.
    This value is superior to that shown in the previous measurement due
    to a higher cavity finesse, the far-detuned driving, and the pre-selection of strongly coupled
    atoms by the transport feedback and a push-out technique (see Supplemental Material).
    The ratio between the rates shown in Figs.~\ref{fig:CavityBackaction}(c)
    and \ref{fig:CavityBackaction}(b) is given by Eq.~\ref{eq:RatesRatio}, 
    and when the resonator and the probe are resonant, a maximum
	cavity collection efficiency of $\beta\!\approx\!90.4\pm 1.4\,\%$ is obtained.  
    \begin{figure}
		\includegraphics[width=1\columnwidth]{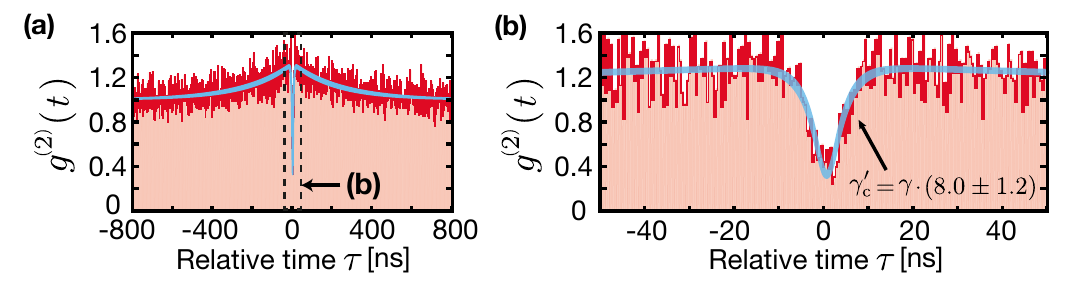}
 		\caption{(a) Autocorrelation function of the cavity output showing the strongly narrowed antibunching dip
        in (b)).
        The curve is a fit to the phenomenological model described in the Supplemental Material.
        The width of the curve represents the one-sigma confidence interval (time bins of $40\,\text{ns}$).}
        {\label{fig:g2Function}}
	\end{figure}

    Due to the presence of a 
    single quantum emitter, the emission collected by the cavity corresponds to a continuous stream of 
    single photons. The quantum nature of the cavity output is confirmed by performing a Hanbury Brown--Twiss
    experiment~\cite{Brown1956} where the field is split in two 
    and detected by independent SPCMs. An antibunching dip is observed in the autocorrelation function
    $g^{(2)}(\tau)$ of the output for $\tau\!=\!0$,
    where $\tau$ represents the delay time between both detector's readouts (see Fig.~\ref{fig:g2Function}
    and the Supplemental Material).
    The measurement yields a minimum of $g^{(2)}(0)\!=\!(0.34\pm 0.05)\!<\!0.5$,
    limited by the 
    detector jitter and consistent with a true single-photon light field.
    The exponential rise of the dip of $(2\gamma_\text{c}^{\prime})^{-1}\!=\!(3.3\pm 0.5)\,\text{ns}$
    is almost an order of magnitude shorter 
    than the natural decay time of the atom ($(2\gamma)^{-1}\!=\!26.24\,\text{ns}$),
    thus providing additional evidence of the clear
    Purcell enhancement in our system, the homogeneous nature of the broadening, and the high repetition
    rate available when using the platform as an SPS.

  In conclusion, we have demonstrated a large Purcell effect in a
  strongly coupled light-matter platform, and shown its performance as an efficient
  and high-bandwidth single-photon interface. By addressing the cycling transition of an
  atom at the center of the cavity mode, the Purcell factor of our system would increase 
  up to $f_\text{P}\!\approx\!190$ (considering a
  total recovery of the cavity finesse). The resulting cavity collection
  efficiency ($99.5\%$) would be of central importance not only for SPS but for any application with
  high detection efficiency requirements (e.g.~\citep{Rosenfeld2017}).
  The high cavity decay rate can also be exploited to bridge the bandwidth
  mismatch between incoming fast photons (\eg from quantum dots~\citep{Zopf2017})
  and long-lived stationary emitters,
  like atoms or ions~\cite{Meyer2015}. This is
  of particular importance for high-bandwidth quantum memories in hybrid links. 
  In this context, it is critical that the coherent interaction between photonic and stationary
  qubits is faster than the cavity decay rate (and ideally higher than the fast-photon bandwidth).
  In our system, this can be achieved by simply
  coupling ensembles instead of single atoms, where collective interaction rates could approach the GHz regime.

    We thank Lothar Ratschbacher and Sutapa Ghosh for their contributions
    to the early stage of the experiment and insightful discussions. This
    work has been financially supported by the Deutsche Forschungsgemeinschaft
	SFB/TR 185 OSCAR, 
	the Bundesministerium f\"ur
    Forschung und Technologie (BMFT, Verbund Q.com-Q),
    and by funds of the European Commission training
    network CCQED and the integrated project SIQS.
    J.\,Gallego, T.\,Macha and M.\,Martinez-Dorantes also acknowledge support
    by the Bonn-Cologne Graduate School of Physics and Astronomy.

\section*{Supplemental Material}

\subsection*{Experimental Methods}\label{app:exp}
\subsubsection*{Finesse degradation}\label{app:finesse}
\begin{figure}
		\includegraphics[width=1\columnwidth]{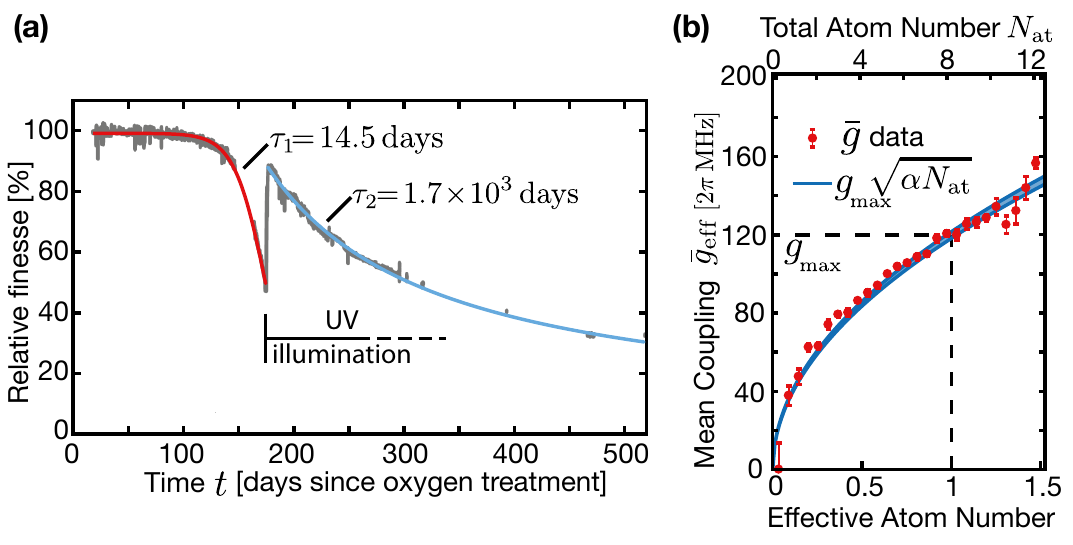}
 		\caption{(a) Decay of the finesse under vacuum after the first oxygen treatment (monitored with
        $770\,\text{nm}$ light) represented as a percentage of the initial value.
        The initial decay corresponds to an exponential increase in losses $\mathcal{L}$ of the form 
        $\mathcal{L}(t)\!=\!\mathcal{L}_0\!+\!\Delta\mathcal{L}\cdot\text{exp}(t/\tau_1)$ (red curve)
		yielding a time constant of $\tau_1\!=\!14.5\,\text{days}$.
        With constant UV illumination the finesse quickly recovers
		(sudden positive slope) and the decay process slows down, following the curve predicted by the oxygen
        depletion model~\cite{Gangloff2015} given by
        $\mathcal{L}(t)\!=\!\mathcal{L}_0\!+\!\Delta\mathcal{L}\cdot(1\!-\!\text{exp}(-t/\tau_2))$
        (blue) with a time
        constant of $\tau_2\!=\!1670 \,\text{days}$. (b) Effective coupling of an atomic ensemble for an
        increasing number of atoms. $N_\text{at}$ is estimated from fluorescence images of the ensemble.
        $\alpha\!=\!0.12$ is the reduction factor accounting for imperfect positioning of the atoms.
 		 }{\label{fig:FinesseAndCollectiveG}}
	\end{figure}

    The initial cavity field decay rate of $\kappa\!=\!2\pi\times (24.5\pm 0.8)\,\text{MHz}$~\cite{Gallego2016}
    increased
    after the resonator was placed under high-vacuum conditions ($\sim\!10^{-9}\,\text{mbar}$),
    due to a sudden 
    rise of optical losses on the fiber-mirrors' coating. This resulted in a finesse degradation
    where the type of decay and the time scales involved are strongly influenced by the presence of ultraviolet (UV)
    light on the mirrors. We observe that UV radiation turned
    a rapid exponential decay of the finesse into a slower
    decline with a half-life of $\sim\!300$ days (see Fig.~\ref{fig:FinesseAndCollectiveG}(a)).
    As a consequence of the finesse variations, the bandwidth of the resonator during
    the measurements presented in the main text varies between two and three times the initial value.
    We observe that the finesse can be recovered (up to $80\%$
    of the initial value) when flushing the vacuum apparatus with pure oxygen.
    The recovery process takes place
    at rates much faster than the ones predicted by the oxygen depletion
    model~\cite{Brandstaetter2013,Gangloff2015}. It remains unclear why 
    UV radiation slows down the degradation process in vacuum, and if it affects both mirrors equally
    (the surface layer of the mirrors coating is composed
    of $\text{SiO}_2$ and $\text{Ta}_2\text{O}_5$ for the LT and HT mirrors, respectively).\\

    \subsubsection*{Effective coupling strength}\label{app:coupling}
    The interaction strength between an atom and the resonator mode varies in our experiments depending
    on two factors: the
    transition addressed by the cavity, and the distance of the atom to the center of the
    Gaussian transversal cavity mode. Regarding the first factor, the strongest coupling is associated
    to the cycling transition $\ket{F,m_F}\!=\!\ket{2,2}\!\leftrightarrow\!\ket{3^\prime,3^\prime}$, corresponding to
    $g_\text{max}\!=\!2\pi\times 120\,\text{MHz}$ for an atom at the center. This is achieved in our
    experiments when using a circularly-polarized probe laser injected into the cavity. After a few
    cycles, the light optically pumps the atoms to the outermost
    Zeeman sublevel of the ground state ($\ket{2,2}$), thus ensuring
    the addressing of the strongest transition.
    The quantization axis is defined by a small bias magnetic field parallel to the resonator and to the electric
    field of the red-detuned dipole traps.
    This is not the case
    when using the lin$\perp$lin side-probe laser, where the resulting atomic steady-state
    population is a
    mixture of different $m_F$-sublevels, with the population distribution
    depending on the frequency of the driving field.
    For instance, in the cavity backaction measurements, the side probe is red-detuned by
    $63\, \text{MHz}$
    from the ac-Stark shifted atomic resonance, to ensure
    one-dimensional polarization-gradient cooling that extends the trapping lifetime.
    This detuning and polarization provides on average a population of
	the outermost $m_F$ sublevels of $40\,\%$,
    as compared to the $10\,\%$ present when pumping at resonance (as the master equation simulations suggest). 
    As a consequence, the atomic excitations do not
    correspond to the closed cycle and the coupling strength associated to an externally driven atom
    is effectively reduced to $g_\text{eff}\!<\!g_\text{max}$. The exact reduction factor is
    extracted from the master equation simulations. A way to address the cycling transition when driving the atom
    with linearly-polarized light would be to substantially increase the trap depth. The resulting strong ac-Stark 
    shift would lift the degeneracy of the $m_F$-sublevels of the excited states, thus allowing for the frequency
    selection of the $\sigma_+$ component of the light. This would correspond to the cycling transition if
    the atom has been 
    previously optically pumped to the ground-state $\ket{2,2}$ sublevel by the cavity probe.

	With respect to the different possible positions of the atom in the cavity mode,
    the typical loading leads to a distribution of coupling
    strengths following a Gaussian curve of average $\bar{g}_\text{eff}\!<\!g_\text{eff}$ and of typical
    width
    $\sigma_g\!\approx\!2\pi\!\times\!18.2\,\text{MHz}$ when the transport
    feedback technique is employed.
    When an ensemble (as opposed to a single atom) is coupled to the resonator, we observe that the 
    coupling scales phenomenologically with the number of atoms as 
    \begin{equation}\label{eq:PositionDependance}
    	\bar{g}_\text{eff}\!\approx\!\sqrt{0.12\cdot N_\text{at}}\,\,g_\text{eff}\, .
    \end{equation}
    where the factor 0.12 accounts for the average distance to the cavity center,
    as shown in Fig.~\ref{fig:FinesseAndCollectiveG}(b).   
	The estimated average coupling from Fig.\ref{fig:Setup}(c) scales slightly more favorable than
    expected from Eq.~\ref{eq:PositionDependance}, due to the transport feedback scheme pre-selecting
    atoms at positions of strong coupling. Additionally, the 2D optical transport capabilities of our system
    could be used to deterministically place single atoms at the optimum position in the resonator (i.e.,
    by monitoring the cavity reflection while performing a 2D position scan).

    \subsubsection*{Number of atoms coupled to the cavity ($N_\text{at}$):
    deterministic transport and push-out sequence.}\label{app:number}
    As introduced in the main text, the influence of the atom on the reflective spectrum of
    the cavity is employed to detect the atom's presence.
    In particular, when an atom enters the cavity, it shifts its resonance and the cavity-probe beam is
    reflected leading to a rise in the SPCM counts, which is then used to stop the transport.
    This signal saturates (i.e., the probe is fully reflected) in the case of strong coupling, meaning
    that the technique does not distinguish between a single strongly coupled atom or an ensemble of them.
    In order to ensure that only one atom is
    coupled to the resonator, we load the conveyor belt sparsely, such that the distance between the 
    few trapped atoms is longer than the transversal cavity mode.
    The reflection threshold used to stop the transport is typically set at $90\%$
    of the maximum of the reflected power.
    
    By performing fluorescence imaging inside the cavity region after the deterministic transport technique, we
    observe that less than $15\%$ of the cases lead to the coupling of more than one atom
    (resulting in an average of $\bar{N}_\text{at}\!=\!1.07$ when also considering cases with zero atoms).
    This sets an upper limit, as the camera
    also detects fluorescence from atoms close to the cavity mode that are not coupled to it.
    In the Purcell-broadening measurements we employ the cavity-probe laser as a 
    side probe (since it allowed for a wider frequency scan),
    and therefore the transport feedback technique is not available. In that case, the average
    number of atoms is estimated by counting the fraction of traces that have one or more atoms coupled to the
    cavity (by detecting photons emitted into the resonator), and assuming that the loading process is
    random and, therefore, Poissonian.
    We found that $77\%$ of the cases yielded a cavity-output signal over the noise floor (signaling the presence
    of atoms in the cavity). The $23\%$ of cases with zero atoms leads to a Poissonian
    distribution with an average of $\bar{N}_\text{at}\!\approx\!1.5$.

    In the context of the cavity backaction measurements,
    the signal-to-noise ratio of a coupled-atom’s scattering into free space
    is reduced in the presence of uncoupled atoms in the proximity of the cavity mode, the fluorescence
    of which is also collected by the high-NA lenses.
    To avoid such contamination of the fluorescence, we
    push the atoms that are not coupled to the resonator out of the optical trap.
    This is done by first using the
    cavity-probe light to only pump the atom in the resonator into the \textit{dark state} manifold
    $\ket{F\!=\!1}$ and, subsequently, performing
    a push-out pulse~\cite{Kuhr2005} with the side-probe beam. This expels all the atoms
    remaining in the \textit{bright state} $\ket{F\!=\!2}$ out of the trap.
    The probability of an atom being pumped to the dark state increases with its coupling strength. 
    The technique leads to a preselection of atoms
    with stronger coupling (manifesting as a rise in the average cooperativity, as shown in the main text),
    since those that are weakly coupled have a higher chance of remaining in the bright state
    and being expelled off the trap. As a consequence, the loading in these measurements is expected to
    be lower than $\bar{N}_\text{at}\!=\!1.07$.

\subsubsection*{Trapping lifetime corrections}\label{app:lifetime}
    Under external laser driving, the atoms continuously scatter photons in all directions resulting in
    emission recoil events that increase the temperature of the trapped atoms.
    The polarization gradient of the driving field (side-probe laser)
    only provides cooling along the illumination
    axis and, therefore, the atoms accumulate thermal motion in the remaining two directions.
    This results in trapping losses that hinder the precise estimation of
    photon emission rates. 
    
    Additionally, the loss rate depends on the resonator detuning
    due to cavity backaction on the total atomic scattering rate.
    Such a dependence needs to be characterized, in order to obtain the atom-loss-corrected
    scattering rates. We average over 150 SPCM data traces for each cavity
    resonance frequency; from these average traces we infer the atom loss rate from the decay in the counts.
    Since different coupling strengths lead to different trapping times, the average decay curve is not
    a single exponential, but a distribution of them. The data is, therefore, fit to a more general
    \textit{stretched exponential}, defined by $A\cdot\text{exp}[(-t/\tau)^\beta]$. This function
    represents the time evolution
    of a system that is driven by a specific distribution of decay processes (given by $\beta$),
    each with a different amplitude $A_i$ and
    lifetime $\tau_i$~\cite{Lee2001}. Although this particular function does not contain
    a full model of the heating
    mechanism, we find that such a phenomenological approach is enough for the scattering rate evaluation.

    The amplitude $A$ of each averaged trace yields a direct estimation of the output
    rate of the cavity $R_c$, as
    it represents the scattering rate before the heating processes take place. In addition, the decay time of
    each curve corresponds to the trapping lifetime for the different cavity detunings.
    The decay behavior is used to estimate the rate $R_\text{f-s}$, which is extrapolated from
    the amount of accumulated 
    photoelectrons $n_\text{ph}$ on the EMCCD chip during a 100-ms-long exposure.
    We assume that the free-space emission follows
    the same decay as the one obtained
    from the corresponding SCPM traces, and consider $R_\text{f-s}$ as the amplitude of the exponential that 
    would lead to $n_\text{ph}$ photoelectrons when integrating over $100\,\text{ms}$. 

\subsubsection*{Single-photon statistics}\label{app:sps}
\begin{figure}
		\includegraphics[width=1\columnwidth]{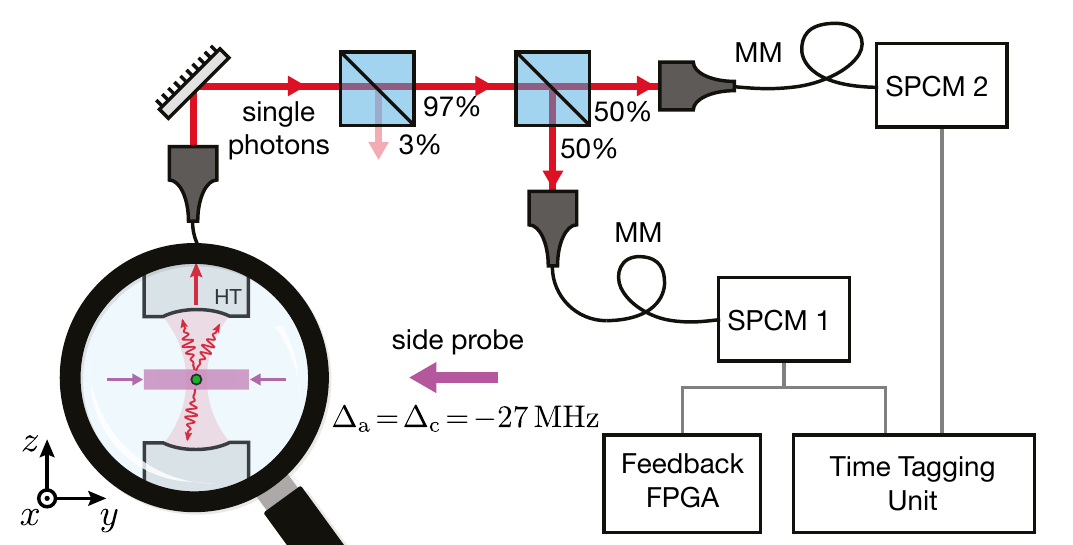}
 		\caption{Experimental setup for the single-photon generation.
        A single atom is driven by the side probe (and a repumper field, not shown), and
        its emission is efficiently collected by the resonator (magnified sketch).
        The light that 
        leaves through the high-transmission (HT) mirror
		is guided through the fiber to a $97\%/3\%$ beam splitter, used to couple the probe into the cavity during
        the feedback transport.
		The single photons are split and sent to both multi-mode (MM-) fiber-inputs of the detectors (SPCM 1,2),
        the output of which is registered in a time tagging unit ($81\,\text{ps}$ resolution).
        The signal from SPCM 1 is duplicated and sent to an FPGA card that computes and generates the signal
        to stop the optical transport.}
        {\label{fig:g2Setup}}
	\end{figure}

	 In order to characterize the quantum character of the light emitted by the coupled system, 
     we perform a simple experiment -- depicted in Fig.~\ref{fig:g2Setup} -- where we externally
     drive a single (coupled) atom and collect its emission with the resonator. The photon statistics 
     of the cavity's output field is then analyzed by performing a Hanbury Brown--Twiss
     experiment~\cite{Brown1956} in which the field is split and subsequently detected by two
     photon counting modules. The cross-correlation between the time-resolved signals of both photodetectors
     yields the second-order intensity correlation function $g^{(2)}$ of the field
    \begin{equation}
		g^{(2)}(\tau) = \frac{\left<c_1(t)\,c_2(t+\tau)\right>}
        {\left< c_1(t)\right >\left< c_2(t)\right >}\, .
	\end{equation}
    Here $c(t)$ is the number of counts detected at time $t$ (either 0 or 1), $\left<\right>$ is the time average
    for sufficiently long periods, $\tau$ is the delay time between both detection traces, 
    and the subindices 1 and 2 represent the photodetectors.
   
    Effects characteristic of single-photon sources, such as antibunching (corresponding to the absence of two
    or more photons emitted at the same time), can be directly observed in the $g^{(2)}$ function.
    In particular, the measured cross-correlation of the cavity output (depicted in Fig.~\ref{fig:g2Function}
    in the main text)
    shows both an antibunching dip at $\tau\!=\!0$, followed by a bunching feature ($g^{(2)}\!>\!1$) around the 
    central dip. While the dip manifests the quantum nature of the field,
    the bunching behavior is attributed to insufficient optical power of the repumper field 
    which causes the atom to spend a considerable fraction of the time in the dark state.
    The typical emission pattern is thus comprised of emission windows separated by “dark”
    periods, and the compression or
    “clustering” of photons in packets leads to the bunching. 
    The behavior can be described phenomenologically by a simple model given by (see e.g.~\cite{Kitson1998})   
    \begin{equation}\label{eq:g2Model}
			g^{(2)}(\tau) = 1-(1+b) e^{-2\tau\gamma_\text{c}^\prime}+be^{-\tau/\tau_\text{b}}\, ,
	\end{equation}
    where $b$ and $\tau_b$ describe the amplitude and decay time that characterize the photon bunching, and
    $(2\gamma_\text{c}^\prime)^{-1}$ stands for the enhanced atomic decay rate.
    
    In this case
    $(2\gamma_\text{c}^\prime)^{-1}\!=\!(3.3\pm 0.5)\,\text{ns}$, which is to be compared to the natural decay time of
    $(2\gamma)^{-1}\!=\!26.24\,\text{ns}$.
    We account for the limited time resolution of the detectors by convolving Eq.~\ref{eq:g2Model}
    with a Gaussian of width $\sigma\!=\!1.35\,\text{ns}$, which describes the specified detectors’ jittering.
    The fast atomic decay and the considerable bunching amplitude ($b=0.33$) reduce the effective width of the
    antibunching dip. In combination with the detectors jittering, the feature is washed out
    and the central value of the fit model rises, leading in our case to the value $g^{(2)}(0)\!=\!0.34\pm 0.05$.
    
\subsection*{Theoretical considerations}\label{app:theo}
\subsubsection*{The driven, dissipative system}\label{app:simplemodel}
    In the absence of dissipative channels, the closed atom--cavity system is fully described
    by the Jaynes--Cummings
    Hamiltonian~\cite{Jaynes1963} which, under the rotating wave approximation, is given by
      \begin{equation}\label{eq:JCHam}
                  \hat{H}_{\text{JC}} = \hbar\omega_\text{a}\,\hat{\sigma}^\dagger \hat{\sigma}
                          + \hbar\omega_\text{c}\,\hat{a}^\dagger \hat{a}
                          + \hbar g \left(\hat{\sigma}^\dagger \hat{a} +
                          \hat{\sigma}\,\hat{a}^\dagger \right)\, ,
      \end{equation}
    where $\hbar$ is Plank's constant and  $\hat{\sigma}$ and $\hat{a}$ are the atomic lowering and the photon
    annihilation operators respectively.
    This simple description is enough to characterize the energy bands of a single excitation
    in the coupled system (shown
    in Fig.~\ref{fig:Setup}(c) in the main text). However it does not contain the external-driving term necessary to
    explore the Purcell and backaction 
    dynamics, which lie at the heart of the experiments presented here. 

    We consider the atom as a two-level system and that the external driving
    ($\propto \Omega$) is
    weak enough to create maximum one excitation. The open, driven system can then be heuristically described by
    a non-Hermitian Hamiltonian (see e.g.~\cite{Lien2016})
      \begin{equation}
              \hat{H}_{\Omega} = \hat{H}_{\text{JC}} -i\hbar(\gamma\hat{\sigma}^\dagger \hat{\sigma} + \kappa 
              \hat{a}^\dagger\hat{a}) +\frac{\Omega}{2}\,(\hat{\sigma}^\dagger\!+ \hat{\sigma})
      \end{equation}
    that includes the irreversible dissipative losses (imaginary term) and the external
    weak pumping (last term).
    Employing Heisenberg's equation on both cavity and atomic amplitude operators ($\hat{a}^\dagger \hat{a}$ and
    $\hat{\sigma}^\dagger \hat{\sigma}$) -- and assuming a steady state scenario -- provides the scattering rate 
    of photons emitted into the cavity and in free space, respectively:

      \begin{subequations}\label{eq:ScatteringRates2}
      \begin{align}
          R_\text{f-s} &= 2\gamma \langle \hat{\sigma}^\dagger\hat{\sigma}\rangle_s 
          \label{eq:ScatteringRateAtom2}\\
          R_\text{c} &= 2 \kappa\langle \hat{a}^\dagger\hat{a}\rangle_s\, ,\label{eq:ScatteringRateCav2}
      \end{align}
      \end{subequations}
    which leads to Eqs.~\ref{eq:ScatteringRates}(a,b) in the main text.
    Notice that $R_\text{c}$ here represents the amount of photons leaving the cavity,
    of which a fraction
    $\eta_\text{HT}\!=\!67\,\%$ (for the initial finesse)
	is collected through the HT-fiber output. Considering the optical path losses and
    the SPCM quantum efficiency, only $\eta_\text{c}\!\approx\!2.1\,\%$
	of the cavity output rate $R_\text{c}$ is finally detected in the cavity backaction measurements (where
    the finesse decay reduced
    $\eta_\text{HT}$ to $18\,\%$).\\
\subsubsection*{Master equation corrections}\label{app:ME}
    The external side-driving of the atom in our measurements is constituted by a lin$\perp$lin
    polarization
    gradient that effectively drives both $\pi$- and $\sigma_\pm$-transitions between the
    $\ket{F\!=\!2}\!\rightarrow\!\ket{F^\prime\!=\!3}$ sublevels (if
    averaging over several atom positions).
    As a consequence, the
    atom cannot be considered a two-level system any longer. Furthermore, in some cases the driving
    powers employed are of the order of 
    the saturation intensity, and the weak excitation approximation does not hold.
    
    To describe the system in such scenario we resort to the master equation (ME) formalism,
    where the system of interest (described by a reduced density
    matrix) evolves according to a Liouvillian operator containing the environmental degrees of freedom.
    Although analytical solutions of the ME are not available for this level of complexity, 
    there are numerical computational approaches~\cite{Tan1999} that provide steady-state solutions.
    The numerical results are used to benchmark the simplified model. Indeed, we observe that
    the simulations qualitatively confirm the system's behavior predicted from Eqs.~\ref{eq:ScatteringRates},
    except for small correction factors on the
    systems main parameters -- namely the
    coupling strength, the cavity bandwidth, the detunings, and the photon collection efficiencies.
    We simulate the measurement conditions and obtain correction factors for
    said parameters by comparing
    both methods. For example, when simulating the red-detuned illumination used in the cavity backaction
    measurements, we observe that the numerically calculated scattering rates closely follow the behavior
    from Eqs.~\ref{eq:ScatteringRates},
    as long as one accounts for a reduction of $5\%$ in $g$ and a rise of $15\%$ in $\kappa$.
    These effects are then included in the fitting model (except for $g$, which is used as a free
    parameter anyway).\\
        \begin{figure}
		\includegraphics[width=1\columnwidth]{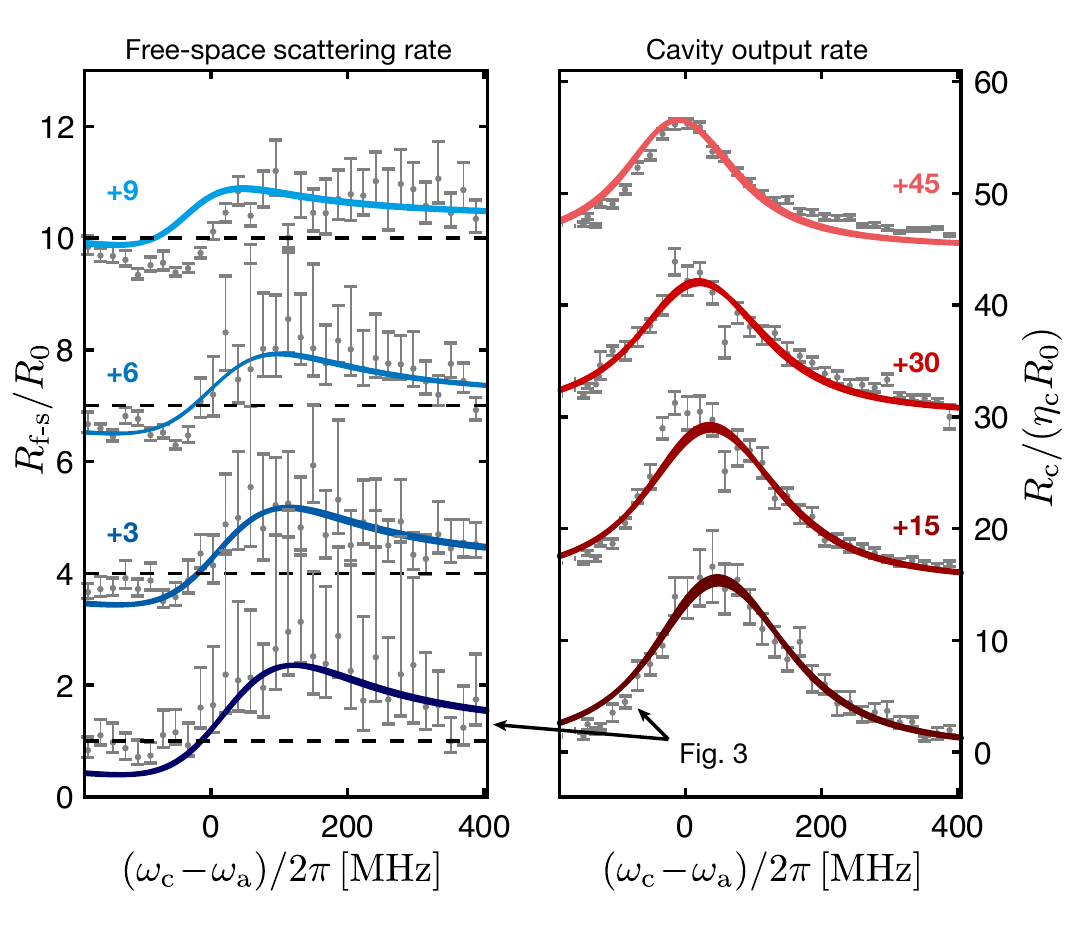}
 		\caption{Free-space (blue, left) and cavity (red, right) emission rates from a single atom for different
        cavity resonance frequencies (horizontal axes). Each plot contains four data sets corresponding to different
        illumination powers ($6.5\,\text{mW}$, $25\,\text{mW}$,
        $65\,\text{mW}$ and $250\,\text{mW}$ from bottom to top), which are vertically
        shifted for better clarity. The curves correspond to a single fit to our model (see discussion and
        Fig.~\ref{fig:CavityBackaction}
        in the main
        text). The dashed, black lines correspond to $R_0$.
 		 }{\label{fig:CavitybackactionFit}}
	\end{figure}

\subsubsection*{Cavity backaction fit}\label{app:FIT}
	The fit shown in Fig.~\ref{fig:CavityBackaction} (main text) is part of a single fit commonly applied to
    four of such pair of curves, which corresponds
    to the same type of measurement with different driving laser powers (see
    Fig.~\ref{fig:CavitybackactionFit}). The fit assumes a single
    free-space collection efficiency $\eta_\text{f-s}$ for all measurements, as well as the same
    resonator bandwidth $\kappa$ and 
    an unknown frequency offset of the cavity resonance  $\omega_\text{c}$. Without such frequency offset,
    we observe an obvious systematic discrepancy between the eight data sets and the common fit.
    The discrepancy
    is considerably reduced when assuming the cavity resonance frequency as a free parameter, while the
    extracted cooperativity only varies by $15\,\%$. The exact origin of the $70\,\text{MHz}$ shift is unclear,
    but we attribute
    it to thermal (or heating) effects on the atoms, which are not included in our simulations. This influences
    the atomic emission properties, \eg different effective ac-Stark shifts for different atom's temperatures.
    Additionally, each measurement pair has the coupling strength
    as a free parameter (four in total), since higher driving powers induce more heating and therefore smaller
    effective coupling constants. This leads to seven free parameters for eight curves.

    The cavity collection efficiency $\eta_\text{c}$ was
    determined in an independent calibration to be $\sim\!2.1\%$.
    A direct measurement of the free-space detection efficiency
    $\eta_\text{f-s}$,
    though, remains challenging.
	The collection capability of the high-NA lens depends on the dipole emission pattern of
    the atom, which varies for different
	cavity resonant frequencies due to optical pumping effects. An approximation considering the emission
    dipole of an atom
    driven with lin$\perp$lin light (and no cavity present) yields $\eta_\text{f-s}\!=\!2.2\,\%$,
    in contrast to the value retrieved
    from the fit of $\eta^\prime_\text{f-s}\!=\!3.3\pm 0.1\,\%$.
    
    The effective cooperativity extracted from the fits differs for the
    free-space and cavity emission curves.
    Although both (cavity and free-space emission) curves are extracted from the same measurement (and therefore
    same average atomic coupling),
    the master equation simulation predicts that optical pumping effects
    lead to a behavior equivalent to reductions on $g$ 
    of $5\,\%$ for $R_\text{c}$ and $11\,\%$ for $R_\text{f-s}$. In combination
    with the fit parameters, this results in
    effective average cooperativities
	of $\bar{C}_{1,\text{c}}\!=\!12.6\pm 0.8$ and $\bar{C}_{1,\text{f-s}}\!=\!10.0\pm 0.6$,
    respectively (their arithmetic mean is the value reported in the main text).

\bibliography{Biblio}
\bibliographystyle{apsrev4-1}
\end{document}